\documentclass[12pt]{article}
\usepackage[pctex32]{graphics}
\begin{document}
\vspace{3cm}
\begin{center}
~
\\
~
{\bf  \Large Entropy of  Black-Branes System and T-Duality}
\vspace{2cm}

                      Wung-Hong Huang\\
                       Department of Physics\\
                       National Cheng Kung University\\
                       Tainan, Taiwan\\

\end{center}
\vspace{2cm}
A general black-branes system under the T-duality transformation will become another smeared system with different dimensional black branes.  We first use  some simple examples to see that both systems have a same value of entropy and then present  a rigorous method to prove this general property.  Using the  property we could easily know the entropy of some complex black-brane systems.

\vspace{6cm}
\begin{flushleft}
E-mail:  whhwung@mail.ncku.edu.tw\\
\end{flushleft}

\newpage
\section{Introduction}
    T-duality is a symmetry of string theory relating small and large distances. It does not shown in point particle theory, indicating that the extended object of strings will experience the spacetime in dramatically different way from that in the  point object of particle.   Especially, it relates different string theories to each other, which preceded the second superstring revolution [1]. 

   T-duality could also relate a $D_p$ brane to another $D_{p\pm 1}$ brane.  The property had been used to find many supergravity backgrounds which are dual to different kind of field theories.   For example, Maldacena and Russo [2] had found the supergravity background which duals to the non-commutative gauge theory through T-duality.  They also discussed the thermodynamics of near-extremal D3-branes with B fields and found that the entropy and other thermodynamic quantities are the same as those of the corresponding D3-branes without B fields.    This means that the thermodynamics of  non-commutative gauge theory is the same as that in the commutative gauge theory [2,3].

Using the T-duality and twist one can construct the supergravity backgrounds which dual to the finite temperature non-commutative dipole field theories [4,5]. The spacetime found for the case of $N=2$ theory is [5]
$$ds_{10}^2 = f(r)^{-1/2}\left[- h(r) dt^2+ dx^2+ dy^2+{ dz^2\over 1+B^2U^2\sin^2\theta}\right]+f(r)^{1/2} \left[h(r)^{-1}dr^2\right.$$
$$+\left. r^2\left(d\theta^2 + \cos^2\theta d\phi^2 +\sin^2\theta(d\chi^2_1+\cos^2\chi_1 d\chi^2_2 + \sin^2\chi_1 d\chi^2_3)\right.\right.$$
$$ \hspace{2.7cm}\left.\left.- {r^2B^2\sin^4\theta \left(\cos^2\chi_1 d\chi_2 + \sin^2\chi_1 d\chi_3\right)^2\over 1+r^2B^2\sin^2\theta}\right)\right],$$
$$B_{z\chi_i}= - {r^2 B\sin^2\theta \left(\cos^2\chi_1 d\chi_2 + \sin^2\chi_1 d\chi_3\right) \over 1+r^2B^2\sin^2\theta}~~~~e^{2\Phi}= {1 \over  1+ r^2B^2\sin^2\theta},~~~
\eqno{(1)}$$
in which $f(r) = 1+{N^4\over r^4}$ and $h(r) = 1-{r_0^4\over r^4}$. Using the resulting metric and dilaton field we can find the entropy S of the solution with a help of symbolic algebra calculation in a computer.  The result shows that it is the same as that without dipole field strength $B$ fields.  The spacetime for the case of $N=1$ and $N=0$ have also been found by us in [5] (which is too lengthy to be cited in here).   After a computer calculation  we also find that the system has a same entropy as that without dipole field strength.  

    It is known that a black-branes system under the T-duality transformation will become another smeared system with different dimensional black branes.  Above property leads us to suspect that both systems have a same value of entropy in the general case and we will prove this property in this short paper. 

  Historically, Suzuki [6] had shown the property in the system of the non-extremal intersecting D-branes.   In this paper we present a more straightforward method to prove this property in a general black-brane system which has non-zero NS-NS B field.
\section{Entropy of  Black-Brane Systems and  T-duality}
To begin with, we quote the formula of T-duality [7].  After the T-duality on $z$ coordinate the metric and dilaton field become: 
$$ \tilde g_{zz} =1/g_{zz}, ~~~\tilde g_{\mu z} = B_{\mu z}/ g_{zz},~~~\tilde g_{\mu \nu} = g_{\mu \nu}-(g_{\mu z}g_{\nu z} - B_{\mu z}B_{\nu z})/g_{zz}\,~~e^{-2\tilde \phi} = g_{zz}~e^{-2\phi} .  \eqno{(2)}$$

{\it Step 1} : First, let us consider the simplest case of black D3 brane which has the metric
$$ds_D^{2} = f(r)^{-1/2}\left[- h(r) dt^2+ dx^2+ dy^2+ dz^2\right]+f(r)^{1/2} \left[h(r)^{-1}dr^2+r^2 d\Omega_5^2\right], \eqno{(3)}$$
in which D=10.   It has zero value of dilaton field.  After the T-duality on $z$ coordinate the metric becomes 
$$ds_{10}^2 = f(r)^{-1/2}\left[- h(r) dt^2+ dx^2+ dy^2\right]+f(r)^{1/2} \left[ dz^2 + h(r)^{-1}dr^2+r^2 d\Omega_5^2\right], e^{-2\phi}= f(r)^{-1/2}, \eqno{(4)}$$
which describes black D2-brane smeared along z-direction.  It is interesting to see that the metric of a black  D0-brane which is uniformly smeared along $x,~y,~z$  is described by [8]
$$ds_{10}^2 = f(r)^{-1/2}\left[- h(r) dt^2\right]+f(r)^{1/2} \left[ dx^2+ dy^2 + dz^2 + h(r)^{-1}dr^2+r^2 d\Omega_5^2\right], e^{-2\phi}= f(r)^{-3/2}. \eqno{(5)}$$
With a T-duality transformation on the directions $x,~y$ we deduce the solution for black D2-brane smeared along the z-direction, which is also described by (4).

 Using the property that the entropy is ${1\over 4}A$ we can calculate the horizon area $A$ from the integration of $\sqrt {|g_{D-2}|}$ in which $|g_{D-2}|$  is the corresponding determinate of the metric without the coordinate $t$ and $r$.   Therefore, in the following we will investigate the invariant property of $|g_{D-2}|$ itself.  Now, it is a easy work to check that the metric in (3), (4)  and (5) have the same value of $|g_{D-2}|$ and thus these systems have a same entropy. {\bf Notice that $|g_K|$ is not an invariant quantity unless $K=D-2$}. 

  How the intrinsic cancelation mechanism, which leads to the invariant property, occurs in the mathematic calculation will be seen in the next example.
\\

   {\it Step 2} : Next, we consider the more general system with the D dimensional background described by the following metric and dilaton field
$$ ds_S^2 = \sum_{i=1}^D A_i dx_i^2, ~~~~e^{-2\phi},~~~~\Rightarrow~~~ds_E^2 = \left(e^{-2\phi}\right)^{2\over D-2}~ds_S^2,\eqno{(6)}$$
in which $ds_S^2$  is the line element in the string frame metric and $ds_E^2$ is that in the Einstein frame.  It follows that 
$$|g_{D-2}| = \left[\left(e^{-2\phi}\right)^{2\over D-2}\right]^{D-2}~\left(\prod_{i=1}^{D-2}A_i\right) =e^{-4\phi}~\left(\prod_{i=1}^{D-2}A_i\right). \eqno{(7)}$$
After T-duality on the $x_1$ coordinate the metric and dilaton field become
$$ d\tilde s_S^2 = \sum_{i \ne1}^D A_i dx_i^2 + {dx_1^2\over A_1}, ~~~~e^{-2\tilde\phi}=A_1~e^{-2\phi},~~~~\Rightarrow~~~d\tilde s_E^2 = \left(A_1~e^{-2\phi}\right)^{2\over D-2}d\tilde s_S^2.\eqno{(8)}$$
It follows that 
$$|\tilde g_{D-2}| = \left[\left(A_1 e^{-2\phi}\right)^{2\over D-2}\right]^{D-2}~\left(\prod_{i\ne1}^{D-2}A_i\right) = \left(A_1^2 e^{-4\phi}\right)~\left({1\over A_1^2}~\prod_{i=1}^{D-2}A_i\right).\eqno{(9)}$$
Above calculations tell us that {\it the extra factor in the dilaton term, i.e. ${A_1^2}$,  just be canceled by the extra factor in the metric term, i.e. ${1\over A_1^2}$}. The intrinsic cancelation mechanism therefore leads to the invariant property of $|g_{D-2}|$ once $|g_{D-2}|$ contains the coordinate which T-duality performs.  {\bf Notice that  $|g_{D-2}|$  is the corresponding determinate of the metric without the coordinates $x_2$ and $x_3$ which can be the {\bf arbitrary coordinates} except $x_1$}. 
\\

{\it Step 3} : Finally, consider the most general black-branes system which is described by the metric $g_{\mu\nu}$ and a dilaton field $\phi$.  After the T-dual transformation on $z$ coordinate we have the new metric $\tilde g_{\mu\nu}$ and a dilaton field $\tilde\phi$. With the help of transformation relation (2) we now have to investigate the following determinate
$$\left|\begin{array} {cccc}
\tilde g_{11}&\tilde g_{12}&\cdot\cdot&\tilde g_{1z}\\
\tilde g_{21}&\tilde g_{22}&\cdot\cdot&\tilde g_{2z}\\
\tilde g_{31}&\tilde g_{32}&\cdot\cdot&\tilde g_{3z}\\
\cdot\cdot&\cdot\cdot&\cdot\cdot&\cdot\cdot\\
\cdot\cdot&\cdot\cdot&\cdot\cdot&\cdot\cdot\\
\tilde g_{1z}&\tilde g_{2z}&\cdot\cdot&\tilde g_{zz}\\
\end{array}
\right| = \left|\begin{array} {cccc}
g_{11}-(g_{1z}g_{1z} - B_{1z}B_{1z})/g_{zz}&g_{12}-(g_{1z}g_{2z} - B_{1z}B_{2z})/g_{zz}&\cdot\cdot&B_{1z}/g_{zz}\\
g_{21}-(g_{2z}g_{1z} - B_{2z}B_{1z})/g_{zz}&g_{22}-(g_{2z}g_{2z} - B_{2z}B_{2z})/g_{zz}&\cdot\cdot&B_{2z}/g_{zz}\\
g_{31}-(g_{3z}g_{1z} - B_{3z}B_{1z})/g_{zz}&g_{32}-(g_{3z}g_{2z} - B_{3z}B_{2z})/g_{zz}&\cdot\cdot&B_{3z}/g_{zz}\\
\cdot\cdot&\cdot\cdot&\cdot\cdot&\cdot\cdot\\
\cdot\cdot&\cdot\cdot&\cdot\cdot&\cdot\cdot\\
B_{1z}/g_{zz}&B_{2z}/g_{zz}&\cdot\cdot&1/g_{zz}\\
\end{array}
\right|
\eqno{(10)}$$
Using the simple property of matrix determinate
$$\left|\begin{array} {ccccc}
a_1+\alpha c_1~~&b_1+\beta c_1~~&\cdot\cdot&\cdot\cdot&c_1\\
a_2+\alpha c_2~~&b_2+\beta c_2~~&\cdot\cdot&\cdot\cdot&c_2\\
a_3+\alpha c_3~~&b_3+\beta c_3~~&\cdot\cdot&\cdot\cdot&c_3\\
\cdot\cdot&\cdot\cdot&\cdot\cdot&\cdot\cdot&\cdot\cdot\\
a_k+\alpha c_k~~&b_k+\beta c_k~~&\cdot\cdot&\cdot\cdot&c_k\\
\end{array}
\right| = 
\left|\begin{array} {ccccc}
a_1&b_1&\cdot\cdot&\cdot\cdot&c_1\\
a_2&b_2&\cdot\cdot&\cdot\cdot&c_2\\
a_3&b_3&\cdot\cdot&\cdot\cdot&c_3\\
\cdot\cdot&\cdot\cdot&\cdot\cdot&\cdot\cdot&\cdot\cdot\\
a_k&b_k&\cdot\cdot&\cdot\cdot&c_k\\
\end{array}
\right|,
\eqno{(11)}$$
relation (10) becomes
$$\left|\begin{array} {cccc}
\tilde g_{11}&\tilde g_{12}&\cdot\cdot&\tilde g_{1z}\\
\tilde g_{21}&\tilde g_{22}&\cdot\cdot&\tilde g_{2z}\\
\tilde g_{31}&\tilde g_{32}&\cdot\cdot&\tilde g_{3z}\\
\cdot\cdot&\cdot\cdot&\cdot\cdot&\cdot\cdot\\
\cdot\cdot&\cdot\cdot&\cdot\cdot&\cdot\cdot\\
\tilde g_{1z}&\tilde g_{2z}&\cdot\cdot&\tilde g_{zz}\\
\end{array}
\right| = \left|\begin{array} {cccc}
g_{11}-g_{1z}g_{1z} /g_{zz}&g_{12}-g_{1z}g_{2z} /g_{zz}&\cdot\cdot&B_{1z}/g_{zz}\\
g_{21}-g_{2z}g_{1z}/g_{zz}&g_{22}-g_{2z}g_{2z}/g_{zz}&\cdot\cdot&B_{2z}/g_{zz}\\
g_{31}-g_{3z}g_{1z}/g_{zz}&g_{32}-g_{3z}g_{2z}/g_{zz}&\cdot\cdot&B_{3z}/g_{zz}\\
\cdot\cdot&\cdot\cdot&\cdot\cdot&\cdot\cdot\\
\cdot\cdot&\cdot\cdot&\cdot\cdot&\cdot\cdot\\
0&0&\cdot\cdot&1/g_{zz}\\
\end{array}
\right|\hspace{3cm}$$
$$= \left|\begin{array} {cccc}
g_{11}-g_{1z}g_{1z} /g_{zz}&g_{12}-g_{1z}g_{2z} /g_{zz}&\cdot\cdot&0\\
g_{21}-g_{2z}g_{1z}/g_{zz}&g_{22}-g_{2z}g_{2z}/g_{zz}&\cdot\cdot&0\\
g_{31}-g_{3z}g_{1z}/g_{zz}&g_{32}-g_{3z}g_{2z}/g_{zz}&\cdot\cdot&0\\
\cdot\cdot&\cdot\cdot&\cdot\cdot&\cdot\cdot\\
\cdot\cdot&\cdot\cdot&\cdot\cdot&\cdot\cdot\\
0&0&\cdot\cdot&1/g_{zz}\\
\end{array}
\right|= \left|\begin{array} {cccc}
g_{11}-g_{1z}g_{1z} /g_{zz}&g_{12}-g_{1z}g_{2z} /g_{zz}&\cdot\cdot&g_{1z}/g_{zz}\\
g_{21}-g_{2z}g_{1z}/g_{zz}&g_{22}-g_{2z}g_{2z}/g_{zz}&\cdot\cdot&g_{2z}/g_{zz}\\
g_{31}-g_{3z}g_{1z}/g_{zz}&g_{32}-g_{3z}g_{2z}/g_{zz}&\cdot\cdot&g_{3z}/g_{zz}\\
\cdot\cdot&\cdot\cdot&\cdot\cdot&\cdot\cdot\\
\cdot\cdot&\cdot\cdot&\cdot\cdot&\cdot\cdot\\
0&0&\cdot\cdot&1/g_{zz}\\
\end{array}
\right|$$
$$= \left|\begin{array} {cccc}
g_{11}&g_{12}&\cdot\cdot&g_{1z}/g_{zz}\\
g_{21}&g_{22}&\cdot\cdot&g_{2z}/g_{zz}\\
g_{31}&g_{32}&\cdot\cdot&g_{3z}/g_{zz}\\
\cdot\cdot&\cdot\cdot&\cdot\cdot&\cdot\cdot\\
\cdot\cdot&\cdot\cdot&\cdot\cdot&\cdot\cdot\\
g_{1z}/g_{zz}&g_{2z}/g_{zz}&\cdot\cdot&1/g_{zz}\\
\end{array}
\right| =
 {1\over g_{zz}^2}\left|\begin{array} {cccc}
g_{11}&g_{12}&\cdot\cdot&g_{1z}\\
g_{21}&g_{22}&\cdot\cdot&g_{2z}\\
g_{31}&g_{32}&\cdot\cdot&g_{3z}\\
\cdot\cdot&\cdot\cdot&\cdot\cdot&\cdot\cdot\\
\cdot\cdot&\cdot\cdot&\cdot\cdot&\cdot\cdot\\
g_{1z}&g_{2z}&\cdot\cdot&g_{zz}\\
\end{array}
\right|. \hspace{6cm}\eqno{(12)}$$
Notice that the above metric determinate transformation is performed in the string frame.  Now, the extra factor ${1\over g_{zz}^2}$ in above result will be canceled by that from the dilation field if we consider the determinate $|g_{D-2}|$ in Einstein frame, as could be easily seen from (9).  Thus we have shown that the entropy of the T-duality transformed black-branes system, which is a smeared higher or lower dimensional black-branes, is the same as the original system.
\section{Discussions}
Let us make following comments to conclude this paper.

1.  As the entropy in statistics is related to the number of the microstate it is reasonable to belive that the number of the microstate does not be changed once the space was performed T-duality transformation. Thus the entropy of the T-duality transformed black-branes system is the same as the original system..  However, this argument could only be {\it directly} applied to the p-branes system such as the Strominger-Vafa model [9], in which the entropy could be calculated {\it directly} by counting the degeneracy of BPS soliton bound states in the p-branes system. 

2.  The steps to construct the Maldacena and Russo [2] model are performing  a T-dual, then a rotation and finally take a T-dual.  As the rotation does not modify the volume the final value of  $|g_{D-2}|$  is therefore invariant.  Thus the entropy is invariant.   When this procedure could be used to found the corresponding supergravity background for arbitrary gauge theory (that in [2] is the $N=4$ theory)  then we may conclude that the arbitrary gauge theory has a same entropy as its corresponding noncommutatitive part of gauge theory.

3. In constructing the dual gravity of noncommutatitive dipole theory [4,5] we perform  a T-dual, then a twist  and finally take a T-dual. As the twist does not modify the volume the entropy is also invariant.  Using the property the entropy associated with (1) could be known without complex computer calculations.

4.  Finally, the property derived in this paper could be the corresponding determinate of the metric without the two arbitrary coordinates, which in evaluating the black brane entropy are the time $t$ and radius $r$ coordinates.  Thus, besides the entropy there are many other invariant quantities under the T-duality transformation. The physical meaning of theses quantities are unknown as yet. 
\\
\\
{\bf Acknowledgments} : The author thanks Kuo-Wei Huang for his encouragement and interesting discussion. 
\\
\\
\begin{center} {\bf  \Large References}\end{center}
\begin{enumerate}
\item J. Polchinski, `String Theory" 1998, Cambridge University Press.
\item J. Maldacena and J. Russo, ``Large N Limit of Non-Commutative Gauge Theories", JHEP 9909 (1999) 025 [hep-th/9908134]; A. Hashimoto and N. Itzhaki, ``Noncommutative Yang-Mills and the AdS/CFT correspondence", Phys. Lett. B465 (1999) 142 [hep-th/9907166].
\item  J.L.F. Barbon, E. Rabinovici, ``On 1/N Corrections to the Entropy of Noncommutative Yang-Mills Theories",  JHEP 9912 (1999) 017 [hep-th/9910019];
T. Harmark and N.A. Obers, `Phase Structure of Non-Commutative Field Theories and Spinning Brane Bound States", JHEP 0003 (2000) 024 [hep-th/9911169]; R. G. Cai. and N. Ohta, ``On the Thermodynamics of Large N Noncommutative Super Yang-Mills Theory", Phys.Rev. D61 (2000) 124012 [hep-th/9910092]. 
\item A. Bergman and O. J. Ganor,``Dipoles, Twists and Noncommutative Gauge Theory," JHEP 0010 (2000) 018 [hep-th/0008030]; A. Bergman, K. Dasgupta, O. J. Ganor, J. L. Karczmarek, and G. Rajesh,``Nonlocal Field Theories and their Gravity Duals," Phys.Rev. D65 (2002) 066005 [hep-th/0103090];  M. Alishahiha and H. Yavartanoo,``Supergravity Description of the Large N Noncommutative Dipole Field Theories," JHEP 0204 (2002) 031 [hep-th/0202131] . 
\item Wung-Hong Huang, `` Thermal Giant Graviton with Non-commutative Dipole Field", JHEP 0711 (2007) 015 [arXiv:0709.0320 ].
\item K. Suzuki, ``Black hole entropy as T-duality invariant,"  Phys.Rev. D58 (1998) 064025 [hep-th/9712224].   
\item T. Buscher, Phys. Lett. B159 (1985) 127; B194 (1987) 59; B201 (1988) 466; E. Bergshoeff, C.M. Hull, T. Ortin,, ``Duality in the Type-II Superstring Effective Action,'' Nucl.Phys. B451 (1995) 547 [hep-th/9504081];  S. F. Hassan, ``T-Duality, Space-time Spinors and R-R Fields in Curved Backgrounds,'' Nucl.Phys. B568 (2000) 145 [hep-th/9907152].
\item  T. Harmark, V. Niarchos, N.A. Obers, ``Instabilities of Near-Extremal Smeared Branes and the Correlated Stability Conjecture", JHEP 0510 (2005) 045 [hep-th/0509011]; ``Instabilities of Black Strings and Branes", Class.Quant.Grav.24 (2007) R1-R90 [hep-th/0701022]. 
\item A. Strominger, C. Vafa, ``Microscopic Origin of the Bekenstein-Hawking Entropy ", Phys.Lett. B379 (1996) 99 [hep-th/9601029 ].  
\end{enumerate}
\end{document}